# Strong Electronic Polarization of the $C_{60}$ Fullerene by the Imidazolium-Based Ionic Liquids: Accurate Insights from Born-Oppenheimer Molecular Dynamics Simulations

Vitaly V. Chaban and Eudes Eterno Fileti

Instituto de Ciência e Tecnologia, Universidade Federal de São Paulo, 12247-014, São José dos Campos, SP, Brazil

**Abstract**. Fullerenes are known to be polarizable due to the strained carbon-carbon bonds and high surface curvature. Electronic polarization of fullerenes is of steady practical importance, since it leads to non-additive interactions and, therefore, to unexpected phenomena. For the first time, hybrid density functional theory (HDFT) powered Born-Oppenheimer molecular dynamics (BOMD) simulations have been conducted to observe electronic polarization and charge transfer phenomena in the $C_{60}$ fullerene at finite temperature (350 K). The non-additive phenomena are fostered by the three selected imidazolium-based room-temperature ionic liquids (RTILs). We conclude that although charge transfer appears nearly negligible in these systems, an electronic polarization is indeed significant leading to a systematically positive effective electrostatic charge on the $C_{60}$ fullerene: +0.14e in [EMIM][Cl], +0.21e in [EMIM][$NO_3$], +0.17e in [EMIM][$PF_6$]. These results are, to certain extent, unexpected providing an inspiration to consider novel $C_{60}$/RTILs systems. HDFT BOMD provides a powerful tool to investigate electronic effects in RTIL and fullerene containing nuclear-electronic systems.

**Introduction**

Fullerenes are nice-looking, highly symmetric chemical formations built exclusively out of carbon atoms. A variety of fullerenes -- from $C_{20}$ to $C_{720}$ -- have been synthesized featuring multiple composition and shapes.[1] Every composition, in turn, gives rise to a plethora of structural isomers.[1] Fullerenes currently find and can potentially find applications in the drug delivery vehicles, for anti-viral activity and anti-oxidant activity.[2-11] Fullerenes are also pursued in the context of solar cells.[12-14] Room-temperature ionic liquids (RTILs) constitute universal solvents,[15] whereas solvation of fullerenes is notoriously problematic due their exclusive structure and interaction behavior.[2, 16-20] If the right solvent for fullerenes were located, this would heavily increase their applications.

Recently, we performed a series of theoretical investigations of $C_{60}$ solvation in the imidazolium-based ionic liquid.[2, 16-18] We also, at least computationally, showed that 1-butyl-3-methylimidazolium tetrafluoroborate helps to disperse $C_{60}$ fullerenes in water.[2] This advance may foster important progress in the biomedical applications of fullerenes. Binding of fullerenes with the imidazolium-based RTILs is not limited to simple van der Waals interaction term, as it was believed before. The accurate binding energy analysis indicated that the ionic liquid perturbs electron density of fullerenes adding certain electrostatic contribution.[16] The recent work proved that moderate changes in the $C_{60}$-RTIL binding lead to a drastic solubility increase even at relatively low temperatures (310-350 K).[16] This phenomenon offers interesting perspectives towards preparation of the concentrated fullerene solutions and calls for a comprehensive additional reconnaissance.

This work reports hybrid density functional theory (HDFT) Born-Oppenheimer molecular dynamics (BOMD) simulations on the $C_{60}$ fullerene and the three selected imidazolium-based RTILs: 1,3-dimethylimidazolium chloride, [MMIM][Cl]; 1,3-dimethylimidazolium nitrate, [MMIM][$NO_3$]; and 1,3-dimethylimidazolium hexafluorophosphate, [MMIM][$PF_6$]. Using an electronic structure approach and finite-temperature dynamics, we provide a detailed analysis of these systems and their electronic properties. We confirm our recent hypothesis that the fullerene-ionic liquid binding force is not exclusively of the van der Waals nature.

**Simulation Methodology**

HDFT BOMD simulation provides an extremely powerful tool to investigate small molecular and ion-molecular systems.[21-23] HDFT to certain extent corrects disadvantages of pure density functionals, such as overestimated electron delocalization, seriously underestimated band gap, overrated binding energies in many important cases, etc.[24] It is frequently said that a good coincidence, when it does occur, between the calculated (using pure functionals) and the experimental band gaps is fortuitous. HDFT also performs better for the self-consistent field (SCF) convergence when the electronic structure case is not trivial. Fullerenes, as most other carbonaceous compounds featuring aromaticity, constitute a complicate convergence case.[16] Our multiple efforts in using pure functionals have resulted in the SCF convergence failure at different time-points during trajectory propagation. Application of the convergence enhancement techniques is not practical for the molecular dynamics simulations due to their heavy computational costs. The outlined problem constitutes one of the practical reasons why BOMD simulations involving fullerenes are still extremely scarce. The study reported here is, in many aspects, unique within the fullerene science.

The gaussian 6-31G basis set supplemented with a polarization function on each non-hydrogen atoms, 6-31G*, has been used to expand the wave function of the system. This basis set was successfully used before to describe electronic structure of the carbonaceous systems.[25] It provides a reasonable balance between accuracy of the properties and computational costs in view of molecular dynamics simulations.

BOMD is an approximation to propagate nuclear equations-of-motion.[26] It is assumed that energy transfer between the atomic nuclei and electrons is absent (although the interaction between them is obviously non-negligible). Therefore, an electronic structure can be recalculated from scratch at every time-point taking into account only the current nuclear positions. The resulting molecular trajectory can be, in such a way, recorded as long as necessary. The integration time-steps in the case of BOMD can be set the same as in the classical molecular dynamics studies. The time-steps can be also enlarged by choosing heavier isotopes of light elements or simulating the system at lower temperature.

The four BOMD systems, as depicted in Figure 1 and summarized in Table 1, have been simulated at 350 K. The temperature was set somewhat higher than the ambient one due to an expectedly slowly conformation dynamics of ionic liquids. All molecular trajectories have been propagated during 10 ps with a time-step of 0.001 ps. That is, 10 000 SCF calculations have been performed for every system of interest.

Electron density delocalization will be discussed in terms of point electrostatic charges assigned through the well-established CHELPG[27] and Hirshfeld[28] procedures. The CHELPG charges are obtained from electrostatic potential (ESP) to reproduce it at the surface of the complex as described by Breneman.[27] The dipole moments will be derived directly from the optimized wave function at each molecular dynamics time-step. The selected molecular orbitals and their spatial localization will be obtained for the geometrically optimized C60+RTIL complexes.

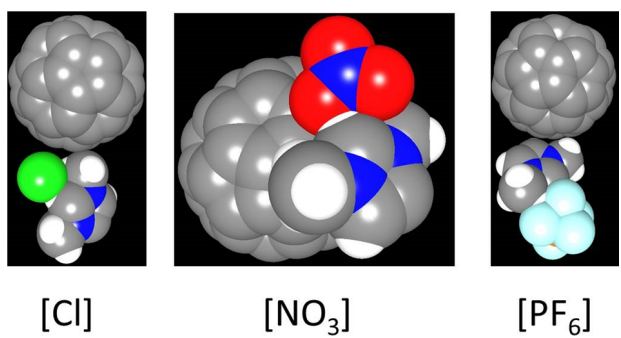

[Cl]    [NO$_3$]    [PF$_6$]

Figure 1. The ion-molecular configurations of [MMIM][Cl], [MMIM][NO$_3$], and [MMIM][PF$_6$] with the C$_{60}$ fullerene after the 10 ps long BOMD simulations at 350 K each. Carbon atoms are gray, nitrogen atoms are blue, hydrogen atoms are white, chlorine atom is green, oxygen atoms are red, fluorine atoms are sky blue, phosphorus atom is orange.

Table 1. The list of the simulated systems and their basic properties. All electrons have been simulated explicitly with no pseudopotentials. Each system was simulated during 10 ps at 350 K with a time-step of 0.001 ps. The resulting trajectories have been processed using the home-made tools

| # | System | # atoms | # electrons | HOMO, eV | LUMO, eV |
|---|--------|---------|-------------|----------|----------|
| 1 | Pristine C$_{60}$ | 60 | 360 | -5.99 | -3.22 |
| 2 | C$_{60}$ + [MMIM][Cl] | 77 | 430 | -5.19 | -3.13 |
| 3 | C$_{60}$ + [MMIM][NO$_3$] | 80 | 444 | -5.54 | -3.18 |
| 4 | C$_{60}$ + [MMIM][PF$_6$] | 83 | 482 | -5.90 | -3.27 |

All electronic structure described above have been conducted using the Gaussian 09 quantum chemistry program suite, revision D (*www.gaussian.com*).[29]

**Results and Discussion**

Figure 2 depicts time evolution of charges. The depicted charge constitutes a sum of all the corresponding charges belonging to the C$_{60}$ fullerene molecule. This sum of charges is expected to be zero for C$_{60}$ and zero for the ion pair. However, the observed differences from zero are significant in the case of the ESP CHELPG charges and insignificant in the case of the Hirshfeld charges. Recall that the Hirshfeld charges were derived from the population analysis and electron density localization, whereas the ESP CHELPG charges were assigned by means of the iterative fitting procedure to reproduce electrostatic potential at the surface of the considered ion-molecular complex. The Hirshfeld charges evolution allows to conclude that electron transfer between the C$_{60}$ fullerene and all RTILs is negligible. Note, however, that only one ion pair was considered instead of bulk ionic liquid. Increase of ion pairs will probably alter the obtained number, but will unlikely change the qualitative conclusion. Interestingly, the C$_{60}$+[MMIM][NO$_3$] system exhibits bizarre fluctuations between 5 and 9 ps. These fluctuations indicate conformational changes in the complex and justify a necessity of the 10 ps long BOMD simulations. Overall, the total Hirshfeld charge on $C_{60}$ in all RTILs is close to zero amounting to -0.002e in [MMIM][Cl], -0.014e in [MMIM][NO$_3$], and +0.029e in [MMIM][PF$_6$]. Here, we provide averages throughout the BOMD trajectory excluding 0-2 ps, which are referred as equilibration). The chloride anion does not foster electron transfer despite its high density of charge. This feature may be viewed unexpected.

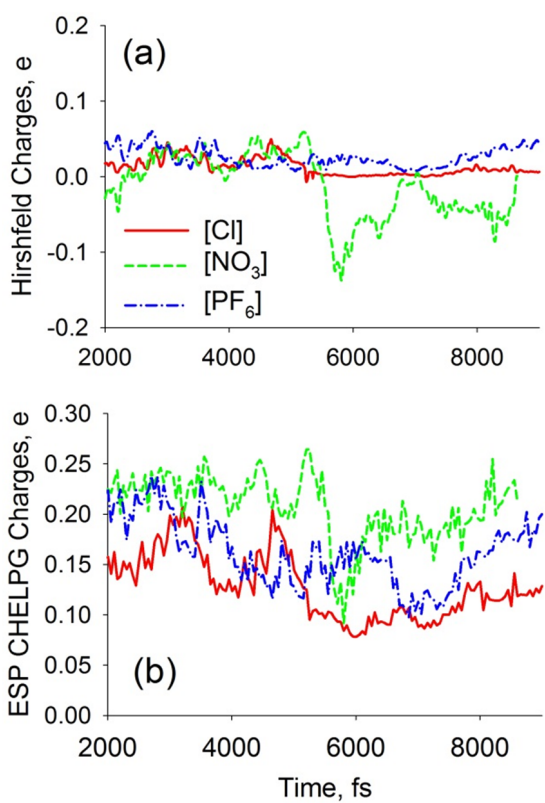

Figure 2. Evolution of the Hirshfeld charges (top) and electrostatic point charges according to the CHELPG scheme (bottom) upon thermal motion of the fullerene and RTIL containing systems: [MMIM][Cl] (red solid line), [MMIM][NO$_3$] (green dashed line), [MMIM][PF$_6$] (blue dash-dotted line). The depicted charges are the total electron charges of the $C_{60}$ fullerene in the course of BOMD.

The ESP CHELPG charges differ from the Hirshfeld charges drastically (Figure 2). The total charge on $C_{60}$ amounts to +0.14e in [EMIM][Cl], +0.21e in [EMIM][NO$_3$], +0.17e in [EMIM][PF$_6$]. Note, the fullerene charges are positive in all RTILs, whereas the Hirshfeld charges were both positive and negative. Remarkably, [EMIM][Cl] appears the least polarizing

RTIL in relation to the $C_{60}$ fullerene, which is contrary to our expectations. The strong polarizing action of $C_{60}$ by ions suggests that interaction between these species goes beyond a simple van der Waals attraction. In addition to the expected π-π stacking between the imidazole ring of the cation and the fullerene surface, the anion plays its own role due to a higher electronegativity of the constituting elements (nitrogen, oxygen, chlorine, fluorine). It is chemically important that our investigation considers a finite-temperature molecular dynamics simulation, in contrast to just optimized geometries.

Figure 3 compares dipole moments in the pristine $C_{60}$ system and upon the RTIL presence. At 350 K, the dipole moment of $C_{60}$ is not zero due to broken symmetry. On the average, it amounts to 0.31 D and fluctuates very significantly. This feature of the fullerenes is a key to understanding of their high polarizability by the ions and polar small molecules. We do not compute a dipole moment of the fullerene separately in its complexes with RTILs, since the total charge on $C_{60}$ deviates from zero in these systems. Thus, the dipole moment cannot be univocally defined in the charged particles.

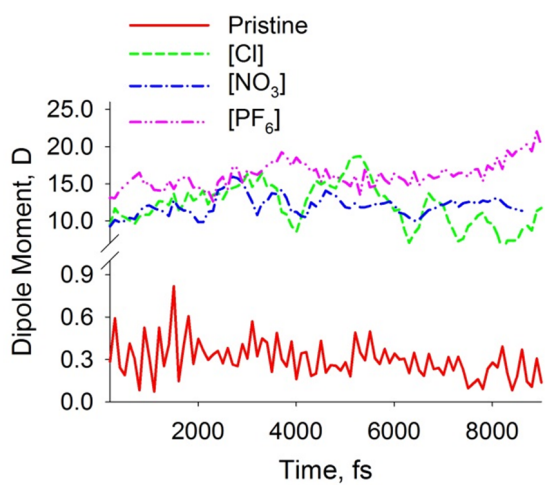

Figure 3. Evolution of the dipole moments during the BOMD simulation of the pristine fullerene $C_{60}$ (red solid line), $C_{60}$ + [MMIM][Cl] (green dashed line), $C_{60}$ + [MMIM][NO$_3$] (blue dash-dotted line) and $C_{60}$ + [MMIM][PF$_6$] (pink dash-dot-dotted line) systems.

The observed peculiarities of the $C_{60}$ electronic properties upon thermal motion foster an interest to the distribution of electron density on the pristine fullerene surface. The electron density anisotropies can be characterized through point charges localized on every carbon atom (Figure 4). These charges were obtained at the end of the BOMD trajectory at 350 K. The system of interest is equilibrated at this point. In the meantime, it is a heated geometry. Whereas the Hirshfeld charges are systematically small (none of them exceeds ±0.02e), the ESP CHELPG charges fluctuate heavily. This feature partially follows from the iterative nature of the procedure of their assignment. Broken symmetry upon thermal motion at 350 K must provide a significant influence on these larger fluctuations. On the other side, fluctuating ESP charges are due to the delocalized valence electron density on the fullerenes. Note that the total charge on $C_{60}$ in this example equals to zero, since this systems contains exclusively the $C_{60}$ molecule.

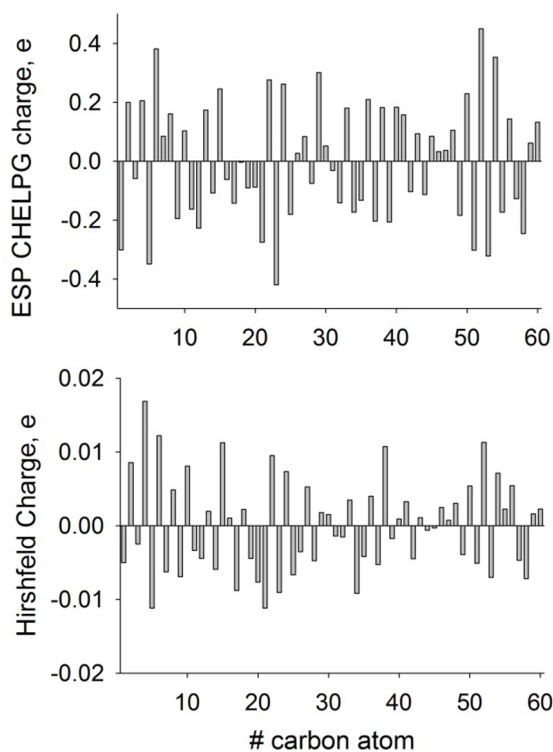

Figure 4. Point electrostatic charges on the pristine fullerene $C_{60}$ at the end of the BOMD simulation at 350 K: (top) calculated according to the CHELPG scheme; (bottom) calculated according to the Hirshfeld scheme. Carbon atom numbering is arbitrary; it does not correspond to the chemical structure of the molecule.

We propose to characterize anisotropy of the electron density distribution by introducing the following simple function of all point charges of the fullerene molecule, $f(q_i) = \sqrt{\sum_i^N q_i^2}$. The sum is taken with respect to all sixty carbon atoms of $C_{60}$. Figure 5 summarizes results on $f(q_i)$ in the case of ESP CHELPG and Hirshfeld atomic charges. The function $f(q_i)$ value in the case of the Hirshfeld charges is poorly dependent on the ion pair. Indeed, [MMIM][Cl] features 0.062e, [MMIM][NO$_3$] features 0.049e, [MMIM][PF$_6$] features 0.048e, whereas pristine $C_{60}$ features 0.049e. Although RTIL may induce local electronic density alterations (i.e. at the site of binding), the overall electron distribution on $C_{60}$ remains virtually unaltered. ESP CHELPG charges exhibit a different trend. The ions rather quench charge fluctuations than promote them.

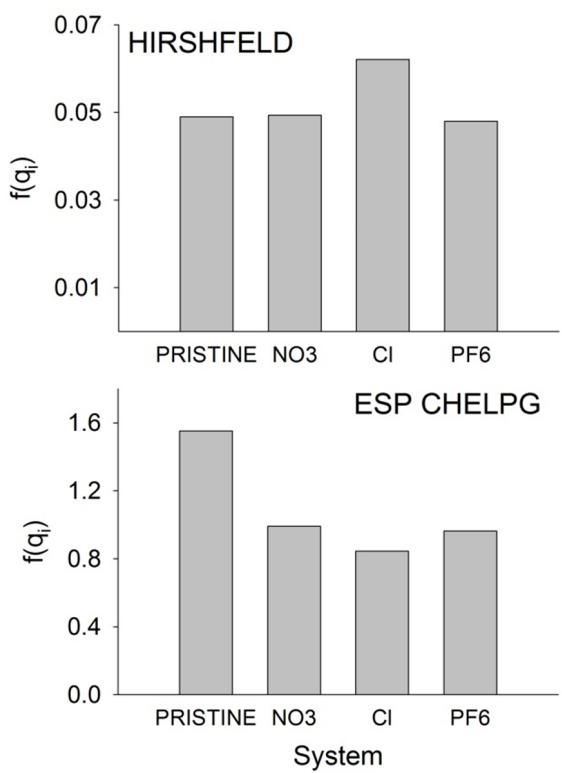

Figure 5. The value of the proposed function, $f(q_i) = \sqrt{\sum_i^N q_i^2}$, to describe an effect of RTIL on the electron density of $C_{60}$.

The highest occupied molecular orbitals (HOMO) and the lowest unoccupied molecular orbitals (LUMO) provide important information not only regarding electrical properties of the system (the band gap, Table 1), but also on the peculiar chemical interactions in it and possible reactivity. The corresponding orbitals are provided in Figure 6. Remarkably, LUMO in all systems is localized exclusively on $C_{60}$. In turn, HOMO is partially shared by $C_{60}$ and RTIL in the case of [MMIM][Cl] and [MMIM][$NO_3$]. However, it is – like LUMO – localized exclusively on $C_{60}$ in the case of [MMIM][$PF_6$]. Figure 7 reports shared molecular orbitals in the valence and conduction bands. Sharing of orbitals between $C_{60}$ and the imidazolium-based RTILs is not common, only a few of them are shared, while a larger fraction of the conduction band bottom arrives from the fullerene and is not shared. LUMO of the chloride anion, E=-3.05 eV, is shared with $C_{60}$ but not shared with the imidazolium cation. In turn, LUMO, E=-0.99 eV, of $MMIM^+$ in the [MMIM][$NO_3$] containing system is partially shared with $C_{60}$. It is located lower along the energy scale than that of nitrate. The similar pattern was observed in [MMIM][$PF_6$] (Figure 7), whereas the $C_{60}$'s density is somewhat shared with the anion in the conduction band, HOMO-5, E = -7.18 eV.

To recapitulate, some valence and conduction band orbitals are shared by the $C_{60}$ fullerene and the ions of RTIL. Some orbitals are even simultaneously shared by $C_{60}$, cation and anion. This is the result of a strong electronic polarizing action exhibited by RTILs. Principal electron energy levels are significantly shifted in the presence of RTIL (Table 1). The band gaps are unfortunately systematically overestimated versus experiment, but this problem is well-known and expected. We do not discuss band gap tuning in this work.

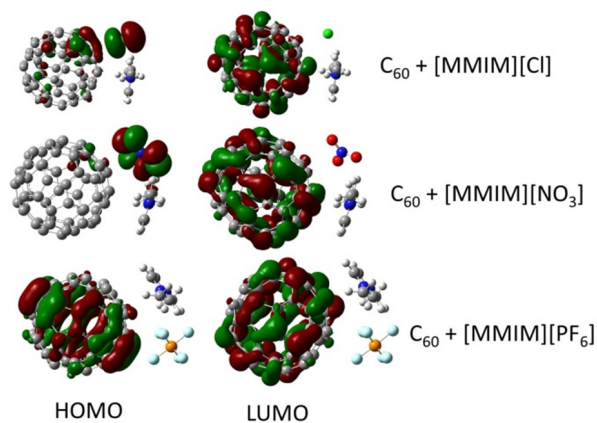

Figure 6. The localization of HOMOs and LUMOs in the simulated systems. The computed and visualized orbitals correspond to the optimized geometries of the complexes, i.e. they do not account for thermal motion at 350 K, as other properties reported in this work. Note that the depicted orbitals are HOMOs and LUMOs of the corresponding entire system rather than of the individual molecules and ions.

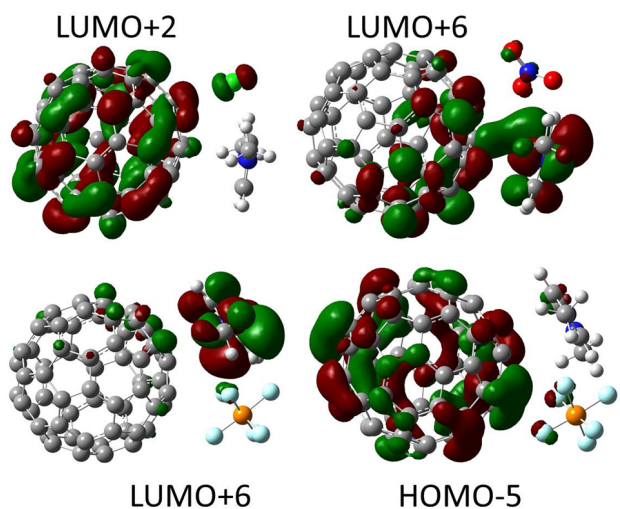

Figure 7. The selected molecular orbitals delocalized between RTIL and $C_{60}$. See figure for designations. Unless originating from the degenerate energy levels, molecular orbital delocalization indicates a strong binding of the corresponding molecular or ionic entities. The computed and visualized orbitals correspond to the optimized geometries of the complexes, i.e. they do not account for thermal motion at 350 K, as other properties reported in this work.

**Conclusions**

For the first time, hybrid density functional theory driven Born-Oppenheimer molecular dynamics simulations have been applied to investigate how the $C_{60}$ fullerene and the three selected imidazolium-based room-temperature ionic liquids interact at finite temperature. By examining the 10 ps long molecular trajectories recorded at 350 K, we conclude that electron transfer between $C_{60}$ and all RTILs is insignificant. However, electronic polarization of the fullerene appears very significant. As a result of just a single RTIL ion pair adsorption at the surface of $C_{60}$, the latter acquires a systematically positive effective electrostatic charge: +0.14e in [EMIM][Cl], +0.21e in [EMIM][NO$_3$], +0.17e in [EMIM][PF$_6$]. The enumerated values are averaged out over the equilibrium parts of the BOMD trajectories. Not only RTIL heavily polarizes $C_{60}$, but it also exhibits a very significant self-polarization upon thermal motion at 350 K. The assigned point charges (average dipole moment equals to 0.31 D) are far from zero suggesting a high reactivity potential.

It must be kept in mind that the current classical molecular dynamics simulations investigating fullerenes in solutions do not account for any of the discovered and outlined specific interactions. Such simulations are unable to provide a highly accurate result and must be employed with certain caution. The existing physical insights must be revisited considering our present findings. Probably, an additional empirical potential effectively strengthening the $C_{60}$-RTIL binding can partially solve the problem of a more adequate description.

The reported results provide an inspiration for searching applications of RTILs and the $C_{60}$ fullerene in combination with one another. One of such applications may be highly efficient solvation of fullerenes in ionic liquids for drug delivery applications and anti-AIDS effects, as we illustrated using large-scale empirical-potential molecular dynamics simulations lately.[2]

**Acknowledgments**

V.V.C. acknowledges research grant from CAPES (Coordenação de Aperfeiçoamento de Pessoal de Nível Superior, Brasil) under "Science Without Borders" program. E.E.F. thanks Brazilian agencies FAPESP and CNPq for support.

**AUTHOR INFORMATION**

E-mail addresses for correspondence: vvchaban@gmail.com (V.V.C.); fileti@gmail.com (E.E.F.)

**REFERENCES**

1. Fowler, P. W.; Manolopoulos, D. E., *An Atlas of Fullerenes*. Dover Publications: New York, 2007.

2. Fileti, E. E.; Chaban, V. V., Imidazolium Ionic Liquid Helps to Disperse Fullerenes in Water. *The Journal of Physical Chemistry Letters* **2014**, *5*, 1795-1800.

3. Fileti, E. E., Atomistic Description of Fullerene-Based Membranes. *J Phys Chem B* **2014**, *118*, 12471-12477.

4. Zhou, X.; Xi, W.; Luo, Y.; Cao, S.; Wei, G., Interactions of a Water-Soluble Fullerene Derivative with Amyloid-Beta Protofibrils: Dynamics, Binding Mechanism, and the Resulting Salt-Bridge Disruption. *J Phys Chem B* **2014**, *118*, 6733-6741.

5. Andreoni, A.; Nardo, L.; Bondani, M.; Zhao, B.; Roberts, J. E., Time-Resolved Fluorescence Studies of Fullerene Derivatives. *J Phys Chem B* **2013**, *117*, 7203-7209.

6. Baoukina, S.; Monticelli, L.; Tieleman, D. P., Interaction of Pristine and Functionalized Carbon Nanotubes with Lipid Membranes. *J Phys Chem B* **2013**, *117*, 12113-12123.

7. Ou, Z.; Jin, H.; Gao, Y.; Li, S.; Li, H.; Li, Y.; Wang, X.; Yang, G., Synthesis and Photophysical Properties of a Supramolecular Host-Guest Assembly Constructed by Fullerenes and Tryptamine Modified Hypocrellin. *J Phys Chem B* **2012**, *116*, 2048-2058.

8. Nedumpully Govindan, P.; Monticelli, L.; Salonen, E., Mechanism of Taq DNA Polymerase Inhibition by Fullerene Derivatives: Insight from Computer Simulations. *J Phys Chem B* **2012**, *116*, 10676-10683.

9. Bianco, A.; Da Ros, T., Chapter 14. Biological Applications of Fullerenes. **2011**, 507-545.

10. Montellano, A.; Da Ros, T.; Bianco, A.; Prato, M., Fullerene C(6)(0) as a Multifunctional System for Drug and Gene Delivery. *Nanoscale* **2011**, *3*, 4035-4041.

11. Calvaresi, M.; Zerbetto, F., Baiting Proteins with C60. *ACS Nano* **2010**, *4*, 2283-2299.

12. Liu, X.; Jeong, K. S.; Williams, B. P.; Vakhshouri, K.; Guo, C.; Han, K.; Gomez, E. D.; Wang, Q.; Asbury, J. B., Tuning the Dielectric Properties of Organic Semiconductors Via Salt Doping. *J Phys Chem B* **2013**, *117*, 15866-15874.

13. Zheng, L.; Liu, J.; Ding, Y.; Han, Y., Morphology Evolution and Structural Transformation of Solution-Processed Methanofullerene Thin Film under Thermal Annealing. *J Phys Chem B* **2011**, *115*, 8071-8077.

14. Hofmann, C. C.; Lindner, S. M.; Ruppert, M.; Hirsch, A.; Haque, S. A.; Thelakkat, M.; Kohler, J., Mutual Interplay of Light Harvesting and Triplet Sensitizing in a Perylene Bisimide Antenna-Fullerene Dyad. *J Phys Chem B* **2010**, *114*, 9148-9156.

15. Hallett, J. P.; Welton, T., Room-Temperature Ionic Liquids: Solvents for Synthesis and Catalysis. 2. *Chem Rev* **2011**, *111*, 3508-3576.

16. Chaban, V. V.; Maciel, C.; Fileti, E. E., Does the Like Dissolves Like Rule Hold for Fullerene and Ionic Liquids? *Journal of Solution Chemistry* **2014**, *43*, 1019-1031.

17.     Fileti, E. E.; Chaban, V. V., Structure and Supersaturation of Highly Concentrated Solutions of Buckyball in 1-Butyl-3-Methylimidazolium Tetrafluoroborate. *J Phys Chem B* **2014,** *118*, 7376-7382.

18.     Chaban, V. V.; Maciel, C.; Fileti, E. E., Solvent Polarity Considerations Are Unable to Describe Fullerene Solvation Behavior. *J Phys Chem B* **2014,** *118*, 3378-3384.

19.     Colherinhas, G.; Fonseca, T. L.; Fileti, E. E., Theoretical Analysis of the Hydration of C60 in Normal and Supercritical Conditions. *Carbon* **2011,** *49*, 187-192.

20.     Maciel, C.; Fileti, E. E.; Rivelino, R., Note on the Free Energy of Transfer of Fullerene C60 Simulated by Using Classical Potentials. *J Phys Chem B* **2009,** *113*, 7045-7048.

21.     Turi, L.; Rossky, P. J., Theoretical Studies of Spectroscopy and Dynamics of Hydrated Electrons. *Chem Rev* **2012,** *112*, 5641-5674.

22.     Rimola, A.; Costa, D.; Sodupe, M.; Lambert, J. F.; Ugliengo, P., Silica Surface Features and Their Role in the Adsorption of Biomolecules: Computational Modeling and Experiments. *Chem Rev* **2013,** *113*, 4216-4313.

23.     Goursot, A.; Mineva, T.; Vasquez-Perez, J. M.; Calaminici, P.; Koster, A. M.; Salahub, D. R., Contribution of High-Energy Conformations to Nmr Chemical Shifts, a Dft-Bomd Study. *Phys Chem Chem Phys* **2013,** *15*, 860-867.

24.     Chaban, V. V.; Prezhdo, V. V.; Prezhdo, O. V., Covalent Linking Greatly Enhances Photoinduced Electron Transfer in Fullerene-Quantum Dot Nanocomposites: Time-Domain Ab Initio Study. *The Journal of Physical Chemistry Letters* **2013,** *4*, 1-6.

25.     Ohba, T.; Chaban, V. V., A Highly Viscous Imidazolium Ionic Liquid inside Carbon Nanotubes. *J Phys Chem B* **2014,** *118*, 6234-6240.

26.     Helgaker, T.; Uggerud, E.; Jensen, H. J. A., Integration of the Classical Equations of Motion on Ab Initio Molecular Potential Energy Surfaces Using Gradients and Hessians: Application to Translational Energy Release Upon Fragmentation. *Chemical Physics Letters* **1990,** *173*, 145-150.

27.     Breneman, C. M.; Wiberg, K. B., Determining Atom-Centered Monopoles from Molecular Electrostatic Potentials - the Need for High Sampling Density in Formamide Conformational-Analysis *Journal Of Computational Chemistry* **1990,** *11*.

28.     Hirshfeld, F. L., Bonded-Atom Fragments for Describing Molecular Charge Densities. *Theoretica Chimica Acta* **1977,** *44*, 129-138.

29.     Frish, M. J.; al., e. *Gaussian 09*, Gaussian, Inc.: Wallingford CT, 2009.